\documentclass[12pt]{article}
\usepackage{epsf}
\usepackage[usenames]{color}
\usepackage{hyperref,epsfig,amsmath,times,amssymb,euscript,graphicx,cite,amssymb,empheq}
\usepackage{wrapfig}
\usepackage{pstricks,fancybox,hyperref}
\usepackage{floatflt}
\fontfamily{pnc}\selectfont

\definecolor{myorange}{rgb}{1,0.5,0}
\begin{document}
\setlength{\parskip}{2ex} \setlength{\parindent}{0em}
\setlength{\baselineskip}{3ex}
\newcommand{\onefigure}[2]{\begin{figure}[htbp]
         \caption{\small #2\label{#1}(#1)}
         \end{figure}}
\newcommand{\onefigurenocap}[1]{\begin{figure}[h]
         \begin{center}\leavevmode\epsfbox{#1.eps}\end{center}
         \end{figure}}
\renewcommand{\onefigure}[2]{\begin{figure}[htbp]
         \begin{center}\leavevmode\epsfbox{#1.eps}\end{center}
         \caption{\small #2\label{#1}}
         \end{figure}}
\newcommand{\comment}[1]{}
\newcommand{\myref}[1]{(\ref{#1})}
\newcommand{\secref}[1]{sec.~\protect\ref{#1}}
\newcommand{\figref}[1]{Fig.~\protect\ref{#1}}
\def\sl2z{SL(2,\Z)}
\newcommand{\be}{\begin{equation}}
\newcommand{\ee}{\end{equation}}
\newcommand{\bea}{\begin{eqnarray}}
\newcommand{\eea}{\end{eqnarray}}
\newcommand{\nn}{\nonumber}
\newcommand{\unit}{1\!\!1}
\newcommand{\mt}{\widehat{t}}
\newcommand{\R}{\bf R}
\newcommand{\X}{{\bf X}}
\newcommand{\T}{{\bf T}}
\newcommand{\PP}{\bf P}
\newcommand{\CC}{\bf C}
\newcommand {\us}[1]{\underline{s_{#1}}}
\newcommand {\uk}{\underline{k}}
\newcommand{\me}{\mathellipsis}
\newdimen\tableauside\tableauside=1.0ex
\newdimen\tableaurule\tableaurule=0.4pt
\newdimen\tableaustep
\def\phantomhrule#1{\hbox{\vbox to0pt{\hrule height\tableaurule width#1\vss}}}
\def\phantomvrule#1{\vbox{\hbox to0pt{\vrule width\tableaurule height#1\hss}}}
\def\sqr{\vbox{%
\phantomhrule\tableaustep
\hbox{\phantomvrule\tableaustep\kern\tableaustep\phantomvrule\tableaustep}%
\hbox{\vbox{\phantomhrule\tableauside}\kern-\tableaurule}}}
\def\squares#1{\hbox{\count0=#1\noindent\loop\sqr
\advance\count0 by-1 \ifnum\count0>0\repeat}}
\def\tableau#1{\vcenter{\offinterlineskip
\tableaustep=\tableauside\advance\tableaustep by-\tableaurule
\kern\normallineskip\hbox
    {\kern\normallineskip\vbox
      {\gettableau#1 0 }%
     \kern\normallineskip\kern\tableaurule}%
  \kern\normallineskip\kern\tableaurule}}
\def\gettableau#1 {\ifnum#1=0\let\next=\null\else
  \squares{#1}\let\next=\gettableau\fi\next}

\tableauside=1.0ex \tableaurule=0.4pt

\bibliographystyle{utphys}
\setcounter{page}{1} \pagestyle{plain}
\numberwithin{equation}{section}

\begin{titlepage}
\begin{center}
 \hfill\\ \vskip 1cm {\sc \large Periodic Schur Process, Cylindric Partitions and $N=2^{*}$ Theory} \vskip 0.5cm
{\sc Amer\,\,Iqbal$^{1,3}$ \,\,\,Can Koz\c{c}az$^{2}$\,\,\, Tanweer Sohail$^{3}$}\\
\vskip 0.5cm
$^{1}${Department of Physics,\\
LUMS School of Science \& Engineering,\\
U-Block, D.H.A, Lahore, 54792, Pakistan.\\} \vskip 0.5cm
$^{2}${ Department of Physics,\\
University of Washington,\\
Seattle, WA, 98195, U.S.A.\\} \vskip 0.5cm
$^{3}$ {Abdus Salam School
of
Mathematical Sciences,\\
G. C. University,\\
Lahore, Pakistan.}
\end{center}
\vskip 2 cm
\begin{abstract}
Type IIA string theory compactified on an elliptic CY3-fold gives rise to $N=2$ $U(1)$ gauge theory with an adjoint hypermultiplet.
We study the refined open and closed topological string partition functions of this geometry using the refined topological vertex. We show that these partition functions, open and closed, are examples of periodic Schur process and are related to the generating function of the cylindric partitions if the K\"ahler parameters are quantized in units of string coupling. The level-rank duality appears as the exchange symmetry of the two K\"ahler parameters of the elliptic CY3-fold.
\end{abstract}
\end{titlepage}

\section{Introduction}
The study of topological strings on Calabi-Yau threefolds has benefited enormously from toric Calabi-Yau threefolds (TCYTs). TCYTs have provided an arena in which various conjectures about the properties of topological strings on Calabi-Yau threefolds can be tested. This has been possible due to the topological vertex formalism which provides a combinatorial way of computing topological string partition function for TCYTs. The combinatorics of the topological string partition functions for TCYTs has been an active area of research for the last few years. Considerable progress has been made in understanding the physical and mathematical reasons for the appearance of various combinatorial structures in these partition functions.

In an earlier paper \cite{Iqbal:2008ra} it was shown that the partition function of $N=2$ $U(1)$ gauge theory with an adjoint hypermultiplet compactified on a circle is given by generating function of the cylindric partitions with a certain profile. The mass of the adjoint hypermultiplet and the coupling constant of gauge theory were related to the profile of the cylindric partitions. In making this identification we used the refined topological vertex \cite{IKV} to calculate the topological string partition function of the geometry which gives rise to the above mentioned gauge theory. In this paper we study the open string partition function by placing a stack of branes in a particular representation. We show that the closed and open topological string partition functions in this case are both examples of periodic Schur process and are related to the generating function of the cylindric partitions once the K\"ahler parameters are quantized in units of string coupling. Since the generating function of cylindric partitions is the character of an irreducible representation of $\widehat{gl}_n$ \cite{tingley} this shows that the topological string partition function can also be interpreted in terms of representations of affine algebra once the K\"ahler parameters are quantized.

The paper is organized as follows. In section 2 we discuss the periodic Schur process and its relation with cylindric partitions. In this section we mainly follow \cite{borodin, tingley}. In section 3 we study the geometry which gives rise to $U(1)$ $N=2^{*}$ theory and we show that the closed and open topological string partition functions for this geometry are examples of periodic Schur process. We also comment on the level rank duality that appears for the cylindric partitions as a manifestation of a $\mathbb{Z}_{2}$ symmetry of the underlying Calabi-Yau threefold.

\section{Periodic Schur Process and Cylindric Partitions}

Periodic Schur process is a random process first studied by Borodin \cite{borodin}. It is a generalization of Schur process first studied by Okounkov and Reshetikhin in \cite{OR}. We will consider the simplest case the periodic Schur process with period one. If we denote the set of all partitions (Young diagrams), including the empty one, by $\mathbb{Y}$ then the periodic Schur process with period one defines a probability measure on $\mathbb{Y}\times \mathbb{Y}$ such that
\bea
P(\lambda,\mu)=\,\varphi_{1}^{|\lambda|}\,\varphi_{2}^{|\mu|}\,s_{\lambda/\mu}({\bf A})\,s_{\lambda/\mu}({\bf B})\times \frac{1}{Z}
\eea
where ${\bf A}$ and ${\bf B}$ are a finite or infinite set of variables and $Z$ is the partition function of the process and is given by
\bea
Z:=\sum_{(\lambda,\mu)\in \mathbb{Y}^2}\,P(\lambda,\mu)
\eea
It was shown in the \cite{borodin} that the above probability measure can also be regarded as a probability measure on cylindric partitions. The sum in the partition function $Z$ above becomes a sum over cylindric partitions. For $\varphi_{2}\mapsto 0$ this becomes ordinary Schur process \cite{OR}. Below we will give a short review of the cylindric partitions beginning with plane partitions, for a more detailed review we refer the reader to \cite{borodin}.

\par{The 3D partitions or plane partitions are generalizations of Young diagrams. They are a weakly decreasing array of non-negative integers $\{\pi_{i,j}\, | \, i,j\geq 1 \}$ }:
\bea
\pi_{i,j}\geq \pi_{i+r,j+s}, \,\,\,\,\,\, r,s > 0.
\eea
They have a very natural pictorial representation in terms of placing $\pi_{i,j}$ boxes at the $(i,j)$ position, similar to a Young diagram $\lambda$ for which we place column of height $\lambda_{i}$ at the $i^{th}$ place. We can also regard a 3D partition as a series of 2D partitions which satisfy the so-called interlacing condition. Let us first define what it means for a 2D partition to interlace another one and give the condition for the partitions obtained from slicing a 3D partition. For two 2D partitions $\mu$ and $\nu$ we say $\mu$ interlaces $\nu$, written as $\mu\succ\nu$, if the heights of the columns of partitions satisfy
\bea
\mu_{1}\geq\nu_{1}\geq\mu_{2}\geq\nu_{2}\geq\mathellipsis\,.
\eea
We slice a 3D partition diagonally by looking at each slice whose projection on the base is given by a set of linear equations parameterized by $a\in {\mathbb Z}: x-y=a$ such that
\bea
\eta(a)=\{ \pi_{i+a,i}\, |\, i\geq max(1,-a+1)\}.
\eea
We have the following condition:
\bea \label{interlace}
\eta(a+1)&\succ&\eta(a)\,,\,\,\,\,\,a<0\,,\\\nn
\eta(a)&\succ& \eta(a+1)\,,\,\,a\geq0\,.
\eea
A skew plane partition is defined as a weakly decreasing array of numbers with the bottom shape $\lambda/\nu=\{(i,j)\in \lambda\,|\,(i,j)\notin \nu\}$, assuming $\lambda\supset\nu$ for $\lambda/\nu$ to make sense. Note that the interlacing condition will depend on the $\nu$ for a skew plane partition with the base $\lambda/\nu$.

\par{The cylindric partitions, first introduced in \cite{GK}, are generalizations of the plane partitions. However, for our purposes, the reparameterization of them in \cite{borodin} is more suitable, which we largely follow. The cylindric partitions naturally appear as the underlying combinatorial structure of the partition function of ${\cal N}=2$ $U(1)$ gauge theory with as single adjoint hypermultiplet as shown in \cite{Iqbal:2008ra}.}

\par
\begin{figure}[h]\begin{center}
\includegraphics[width=3in]{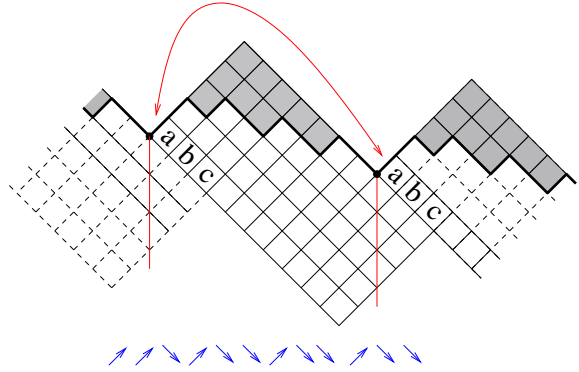}
\caption{\small An example for a cylindric partition: $\lambda=(5,5,\mathellipsis,5)$ and $\nu=(3,2,2,1,1)$ with the shift $d=7$. The grey shaded region shows the excised region.} \label{cyl}
\end{center}\end{figure}

{A cylindric partition is a plane partition with additional symmetry imposed. We will take any 2D partition $\lambda$ of fixed $\lambda_{1}$ such that $\lambda$ is big enough to include another 2D partition $\mu$\footnote{Our notation differs from most of the mathematics literature in the pictorial representation of the 2D partition. They are related by taking the transpose.}. In addition to the conditions defining a plane partition we want require periodicity. To this end, if we take the transposed partition $\lambda^{t}$ and place it on top of the first column of itself after a shift by $d$ boxes, the new plane partition should still be a plane partition as in \figref{cyl}. As indicated in this figure, this modification is equivalent to the requirement that the partitions denoted by vertical red slices are identical, \textit{i.e.}, we can write these types of planes partitions on a cylinder. }

The profile of the cylindric partition is its shape on the cylinder. It can be described by two sets of $N$ numbers where $N$ is the period of the partition \cite{borodin}. It is easy to see that the period is equal to the sum of $\lambda_{1}$ and the shift $d$.  Let ${\bf A}=(a_{1},a_{2},\cdots,a_{N})$ and ${\bf B}=(b_{1},b_{2},\cdots,b_{N})$ such that
\bea
a_{i},b_{i}\in \{0,1\}\,,\,\,\,\,\,a_{i}+b_{i}=1 \,\,\,\,\mbox{for all} \,\,\,\,\,i\in \{1,2,\cdots,N\}
\eea
The profile of the partition is then given by ${\bf A}$ and ${\bf B}$ such that \cite{tingley}
\bea
a_i&=&\left\{
      \begin{array}{ll}
        1, & \hbox{if the boundary slopes up and to the right on the $i$-th diagonal;} \\
        0, & \hbox{Otherwise.}
      \end{array}
    \right.\\\nn
b_i&=&\left\{
      \begin{array}{ll}
        1, & \hbox{if the boundary slopes down and to the right on the $i$-th diagonal;} \\
        0, & \hbox{Otherwise.}
      \end{array}
    \right.
\eea
If we define $n$ and $\ell$ such that $\sum_{i=j}^{N}a_{j}=n$ and $\sum_{i=1}^{N}b_{i}=\ell$ then each profile is equivalent to a partition  $\nu$ such that $\ell(\nu)\leq \ell$ and $\ell(\nu^t)\leq n$. Thus the partition can fit inside a $n\times \ell$ rectangle\footnote{Note that $\lambda_{1}=n$ and $d=\ell$}. This gives the shape of the boundary of the partition near the cut of the cylinder. For the example shown in \figref{cyl} the profile is given by ${\bf A}=(1,1,0,1,0,0,1,0,0,1,0,0)$ and ${\bf B}=(0,0,1,0,1,1,0,1,1,0,1,1)$. As mentioned before, the interlacing condition is modified for a skew plane partition. There is a very direct relation to the sets ${\bf A}$ and ${\bf B}$. Let us first write the interlacing condition:

\bea\nn
\eta{(5)}\prec\eta{(4)}\prec\eta{(3)}&\succ&\eta{(2)}\prec\eta{(1)}\succ\eta{(0)}\\ \nn &\succ&\eta{(-1)}\prec\eta{(-2)}\succ\eta{(-3)}\succ\eta{(-4)}\prec\eta{(-5)}\succ\eta{(-6)}\succ\eta{(-7)}\,,
\eea

where we need to keep in mind the periodicity condition we imposed, $\eta(5)=\eta(-7)$. A quick comparison shows that the 1's in the set $\bf A$ correspond to $``\prec"$ and 0's to $``\succ"$, and vice versa for the set $\bf B$.

For our discussion of $U(1)$ theory with adjoint matter we will choose $\lambda$ to be a rectangular 2D partition of size $n\times \ell$ so that $N=n+\ell$. For $\nu$ trivial the profile is given by \bea\nn
{\bf A}&=&(\underbrace{1,1,\cdots,1}_{n},\underbrace{0,0,\cdots,0}_{\ell}),\\\nn
{\bf B}&=&(\underbrace{0,0,\cdots,0}_{n},\underbrace{1,1,\cdots,1}_{\ell})\,.
\label{profile}
\eea
In terms of the partition $\nu$ the set ${\bf A}$ and ${\bf B}$ are
\bea\label{map}
a_{k}&=&1 \,\,\,\,\mbox{for}\,\,\,\, k=\nu_{j}^{t}-j+1+n\,,\,\,\,j=1,\mathellipsis, n\\\nn
b_{k}&=&1 \,\,\,\,\mbox{for}\,\,\,\,k=n-\nu_{i}+i\,,\,\,\,i=1,\mathellipsis,\ell\,.
\eea

Let us define $\mathbb{G}^{\ell,n}_{\nu}(s)$ to be the generating function of cylindric plane partitions,
\bea
\mathbb{G}^{\ell,n}_{\nu}(s)=\sum_{\mbox{\tiny cylindric partitions $\pi$ of profile $\nu$}}\,s^{|\pi|}\,,
\eea
where $\nu$ is such that $\ell(\nu)\leq \ell$ and $\ell(\nu^t)\leq n$. The above generating function was determined in \cite{borodin} and is given by
\bea
\mathbb{G}^{\ell,n}_{\nu}(s)=\prod_{k=1}^{\infty}\Big((1-s^{kN})^{-1}\prod_{k_{1},k_{2},a_{k_{1}}=1,b_{k_{2}}=1}(1-s^{(k_{1}-k_{2})(N)+(k-1)N})^{-1}\Big),
\eea
where
\bea
(k_{1}-k_{2})(N)=\left\{
           \begin{array}{ll}
             k_{1}-k_{2}, & \hbox{if}\,\,k_{1}-k_{2}>0 \\
             k_{1}-k_{2}+N, & \hbox{if}\,\,\,k_{1}-k_{2}<0\,.
           \end{array}
         \right.
\eea
Using Eq. \ref{map} we can write $\mathbb{G}^{\ell,n}_{\nu}(s)$ as
\bea
\mathbb{G}^{\ell,n}_{\nu}(s)&=&\prod_{k=1}^{\infty}\Big((1-s^{k\,N})^{-1}\prod_{(i,j)\in \nu}(1-s^{h_{\nu}(i,j)+(k-1)N})^{-1}\prod_{(i,j)\notin \nu}(1-s^{h_{\nu}(i,j)+k\,N})^{-1}\Big)\,,\\\nn
&=&\prod_{(i,j)\in \nu}(1-s^{h_{\nu}(i,j)})^{-1}\prod_{k=1}^{\infty}\Big((1-s^{k\,N})^{-1}\prod_{i=1,j=1}^{\ell,n}(1-s^{h_{\nu}(i,j)+k\,N})^{-1}\Big),
\eea
where $h_{\nu}(i,j)$ is the hook length given by
\bea
h_{\nu}(i,j)&=&\nu_{i}+\nu_{j}^{t}-i-j+1\,.\\\nn
\eea

\section{$N=2^{*}$ Theory, Elliptic Calabi-Yau Threefold and Branes}

\par
\begin{figure}[h]\begin{center}
\includegraphics[width=2.5in]{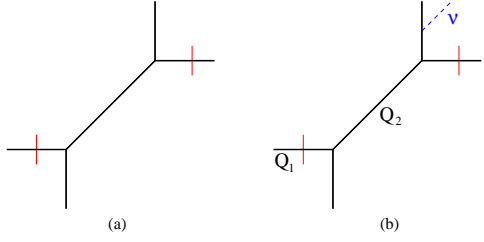}
\caption{\small The web diagram of $X_{H}$ (a) and web diagram of $X_{H}$ with branes (b).}  \label{Xh}
\end{center}\end{figure}

{The ${\cal N}=2$ abelian supersymmetric theory with one massive adjoint hypermultiplet (the $N=2^{*}$) theory can be engineered in type IIA string theory either by using NS5-branes and D4-branes  or by compactfication on a non-compact elliptic Calabi-Yau threefold $X_{H}$ \cite{Hollowood:2003cv}. These two descriptions of the theory are closely related to each other since dualities relate the NS5-brane/D4-brane picture to the elliptic Calabi-Yau. This elliptic Calabi-Yau threefold is an elliptic fibration  and can also be obtained by partial compactification of the resolved conifold, ${\cal O}(-1)\oplus{\cal O}(-1)\mapsto {\mathbb P}^{1}$ \cite{Morrison:1996na}. In the toric/web picture this partial compactification is achieved by identification of the two external legs as shown in \figref{Xh}. In this section we will study the topological string partition function of $X_{H}$ with and without a stack of branes, in the fiber over the elliptic curve.}

\par{It was shown in \cite{Hollowood:2003cv} that the partition function of the abelian theory with the adjoint hypermultiplet can be computed using the topological vertex \cite{Iqbal:2002we,AKMV}. Recently, the corresponding more refined partition function\footnote{The refined partition function of the ${\cal N}=2$ gauge theory is a function of the equivariant parameters $\epsilon_{1}$ and $\epsilon_{2}$ corresponding to the $U(1)\times U(1)$ action on $\mathbb{C}^{2}$  $(z_{1},z_{2})\mapsto (e^{i\epsilon_{1}}\,z_{1},e^{i\epsilon_{2}}\,z_{2})$. The usual partition function corresponds to $\epsilon_{1}+\epsilon_{2}=0$. } was computed using the refined topological vertex \cite{IKV} and shown to be exactly the same as the refined generating function of cylindric partitions of a special profile. The parameters defining this special profile are identified with the K\"{a}hler parameters of $X_{H}$ and on the gauge theory side with the mass of the adjoint and the coupling constant. For a detailed map between the K\"ahler parameters and the gauge theory parameters we refer the reader to \cite{Iqbal:2008ra}. In the geometry considered in \cite{Iqbal:2008ra} there were no branes present. Later in this section we are going to argue that the correspondence between the cylindric partitions and the open string partition function of $X_{H}$ also holds.}

\par{The refined partition function for this geometry can be easily computed using the refined topological vertex. The refined topological vertex has a preferred direction \cite{IKV} and we need to make a choice for it. For simplicity, we glue the vertices along the un-preferred direction and put the branes with representation $\nu$ along the preferred one as shown in \figref{Xh}(b).}\footnote{$ C_{\lambda\,\mu\,\nu}(t,q)=(\frac{q}{t})^{\frac{\Arrowvert\mu\Arrowvert^2+\Arrowvert\nu\Arrowvert^2}{2}}\,t^{\frac{\kappa(\mu)}{2}}\,P_{\nu^t}(t^{-\rho};q,t)\sum
(\frac{q}{t})^{\frac{|\eta|+|\lambda|-|\mu|}{2}}\,s_{\lambda^t/\mu}(t^{-\rho}\,q^{-\nu})\,s_{\mu/\eta}(t^{-\nu^t}\,q^{-\rho})\\
\rho=\{-\frac{1}{2},-\frac{3}{2},\cdots\}\\
\Arrowvert\nu\Arrowvert^2=\sum_{i}\nu_{i}^2\,,\,\,\,\,\kappa(\nu)=\Arrowvert\nu\Arrowvert^2-\Arrowvert\nu^t\Arrowvert^2$.}
\bea\nn
Z_{\nu}(Q_{1},Q_{2},t,q)&=&\sum_{\lambda,\mu}(-Q_{1})^{|\lambda|}(-Q_{2})^{|\mu|}C_{\lambda\mu\nu}(t^{-1},q^{-1})
C_{\lambda^{t}\mu^{t}\emptyset}(q^{-1},t^{-1})\\
\nonumber
&=&q^{-\frac{||\nu||^2}{2}}\widetilde{Z}_{\nu}(t^{-1},q^{-1})\sum_{\lambda,\mu,\eta_{1},\eta_{2}}Q_{1}^{|\lambda|}Q_{2}^{|\mu|}
\left(\frac{t}{q}\right)^{\frac{|\eta_{1}|-|\eta_{2}|}{2}}
s_{\lambda^{t}/\eta_{1}}(t^{\rho}\,q^{\nu})s_{\mu/\eta_{1}}(t^{\nu^{t}}\,q^{\rho})\\\nn
&\times&s_{\lambda^{t}/\eta_{2}^{t}}(q^{-\rho})s_{\mu/\eta_{2}^{t}}(t^{-\rho})\\\nn
&=&M(t,q)^{-1}\,q^{-\frac{||\nu||^2}{2}}\widetilde{Z}_{\nu}(t^{-1},q^{-1})\\
&\times& \sum_{\lambda,\mu}\,Q_{\bullet}^{|\lambda|}\,
s_{\lambda/\mu}\Big(t^{\rho-\frac{1}{2}}\,q^{\nu+\frac{1}{2}},Q_{1}^{-1}t^{-\rho}\Big)\,
s_{\lambda/\mu}\Big(q^{\rho}\,t^{\nu^{t}},\sqrt{\frac{t}{q}}\,Q_{2}^{-1}q^{-\rho}\Big),
\label{final}
\eea
where $M(t,q)=\prod_{i,j=1}^{\infty}(1-t^{-i+1}q^{-j})^{-1}$ is a refinement of the MacMahon function, $t_{1,2}=-\mbox{log}\,Q_{1,2}$ are the K\"ahler parameters of $X_{H}$ and $Q_{\bullet}=Q_{1}Q_{2}$. $\widetilde{Z}_{\nu}(t,q)$ is related to the Macdonald function $P_{\nu}(\mathbf{x};q,t)$,
\bea
P_{\nu^{t}}(t^{-\rho};q,t)&=&t^{\frac{\Arrowvert\nu\Arrowvert^{2}}{2}}\widetilde{Z}_{\nu}(t,q) \\ \nn
&=&t^{\frac{\Arrowvert\nu\Arrowvert^{2}}{2}}\prod_{(i,j)\in\nu}\left( 1-t^{a(i,j)+1}q^{\ell(i,j)}\right)^{-1}, \,\,\,\, a(i,j)=\nu_{j}^{t}-i,\, \ell(i,j)=\nu_{i}-j.
\eea
Using the following identity
\bea
\sum_{\lambda,\mu}\rho^{|\lambda|}s_{\lambda/\mu}(x)s_{\lambda/\mu}(y)=\prod_{k=1}^{\infty}\left(1-\rho^{k}\right)^{-1}\prod_{i,j=1}^{\infty}\left(1-\rho^{k}x_{i}y_{j} \right )^{-1}\,,
\eea
we get\footnote{After analytic continuation to $t^{-1}$ and $q^{-1}$.}
\begin{align}
&Z_{\nu}(Q_{1},Q_{2},t,q)=\,q^{-\frac{\Arrowvert\nu\Arrowvert^2}{2}}\widetilde{Z}_{\nu}(t^{-1},q^{-1})\\\nn
&\times\prod_{k=1}^{\infty}\Big[(1-Q_{\bullet}^k)^{-1}\prod_{i,j=1}^{\infty}
\frac{(1-Q_{\bullet}^{k}Q_{1}^{-1}t^{\nu^{t}_{j}-i+\frac{1}{2}}q^{-j+\frac{1}{2}})(1-Q_{\bullet}^{k}Q_{2}^{-1}q^{\nu_{i}-j+\frac{1}{2}}t^{-i+\frac{1}{2}})}
{(1-Q_{\bullet}^{k}t^{\nu^{t}_{j}-i}q^{\nu_{i}-j+1})(1-Q_{\bullet}^{k}q^{-j}t^{-i+1})}\Big].
\end{align}
The closed string partition function is obtained by taking $\nu=\emptyset$,
\bea\label{cspf}
Z_{\emptyset}(Q_{1},Q_{2},t,q)=\prod_{k=1}^{\infty}\Big[(1-Q_{\bullet}^k)^{-1}\prod_{i,j=1}^{\infty}\frac{(1-Q_{\bullet}^{k}Q_{1}^{-1}t^{-i+\frac{1}{2}}q^{-j+\frac{1}{2}})
(1-Q_{\bullet}^{k}Q_{2}^{-1}q^{-j+\frac{1}{2}}t^{-i+\frac{1}{2}})}
{(1-Q_{\bullet}^{k}t^{-i}q^{-j+1})(1-Q_{\bullet}^{k}q^{-j}t^{-i+1})}\Big].
\eea
The above is also the partition function of the 5D compactified $U(1)$ $N=2^{*}$ theory and can be written as
\bea
Z_{\emptyset}(Q_{1},Q_{2},t,q)&=&\prod_{i,j}(1-Q_{1}\,t^{i-\frac{1}{2}}\,q^{j-\frac{1}{2}})\sum_{n=0}^{\infty}Q_{2}^{n}\,f_{n}(Q_{1},t,q)
\eea
Where the first product on the R.H.S above is the perturbative contribution to the gauge theory partition function and the second factor (the sum) is the non-perturbative instanton contribution and can be written as the chi-y genus of the Hilbert scheme of $\mathbb{C}^2$ \cite{Hollowood:2003cv}:
\bea
f_{n}(Q,t,q)=\chi_{Q}(H_{n})=\int_{H_{n}}\,\mbox{ch}\Lambda_{-Q}(T^{*}H_{n})\,T(H_{n})=\int_{H_{n}}\prod_{j=1}^{dimH_{n}}(1-Q\,e^{-x_{j}})\frac{x_{j}}{1-e^{-x_{j}}}
\eea
where $H_{n}=\mbox{Hilb}^{n}[\mathbb{C}^{2}]$, $T(H_{n})$ is the Todd class of $H_{n}$, $\{x_{1},x_{2},\cdots,x_{dim\,H_{n}}\}$ denote the formal Chern roots of $T(H_{n})$.

 The above partition function can also be obtained directly using the refined Gopakumar-Vafa (GV) invariants \cite{Gopakumar:1998ii} of the curves in this geometry . Recall that the topological string partition function in terms of refined GV invariants is given by \cite{Hollowood:2003cv}
\bea\label{pf}
Z&=&\prod_{\beta\in H_{2}(X,\mathbb{Z})}Z_{\beta}\\\nn
Z_{\beta}&:=&\prod_{j_{L},j_{R}}\prod_{k_{L}=-j_{L}}^{+j_{L}}\prod_{k_{R}=-j_{R}}^{+j_{R}}
\prod_{m_{1},m_{2}=1}^{\infty}
\Big(1-q_{1}^{k_{L}+k_{R}+m_{1}-\frac{1}{2}}\,q_{2}^{k_{L}-k_{R}+m_{2}-\frac{1}{2}}Q_{\beta}\Big)^{(-1)^{2(j_{L}+j_{R})}N_{\beta}^{j_{L},j_{R}}},
\eea
where $Q_{\beta}=e^{-\int_{\beta}\omega}$ is given by the complexified K\"ahler form $\omega$ and $N_{\beta}^{j_{L},j_{R}}$ are the refined GV invariants which are the number of cohomology classes, with spin $(j_{L},j_{R})$, of the moduli space of D2-branes wrapped on the holomorphic curve in the class $\beta$. The moduli space of the D2-brane is not just the moduli space of the holomorphic curve on which it is wrapped because the D2-brane has a $U(1)$ gauge field living on its worldvolume and therefore the moduli space of D2-brane includes the moduli of the flat connections on the curve coming from the gauge field as well as the moduli of the curve. The moduli space of the D2-brane is therefore a $T^{2g}$ fibration over the moduli space of the curve since the moduli of the flat connection over a smooth genus $g$ curve is $T^{2g}$. Since the moduli space of the holomorphic curve in a Calabi-Yau threefold is a K\"ahler manifold, the total moduli space of D2-brane is also K\"ahler manifold such that the Lefshetz action by the K\"ahler class is the diagonal of the $SU(2)_{L}\times SU(2)_{R}$ action on the moduli space. The $SU(2)_{L}$ acts on the fiber and the $SU(2)_{R}$ acts on the moduli space of curve, the base.

In the example we are considering $H_{2}(X_{H})$ is two dimensional generated by the two genus zero curves $C_{1}$ and $C_{2}$. The curve $E:=C_{1}+C_{2}$ has genus one. The holomorphic curves in this geometry are given by (Appendix A)
\bea
(k+1)\,E\,,\,\,\,k\,E+C_{1},\,\,\,\,k\,E+C_{2}\,,\,\,\,\,\,k\geq 0\,.
\label{holo}
\eea
The curves $k\,E+C_{1}$ and $k\,E+C_{2}$ are of genus zero and rigid therefore the moduli space of D2-brane wrapped on these curves is just a point,
\bea
N_{kE+C_{1}}^{j_{L},j_{R}}=N_{kE+C_{2}}^{j_{L},j_{R}}=\delta_{j_{L},0}\delta_{j_{R},0}\,.
\eea
The curve $(k+1)E$ is of genus one but is also rigid therefore the only moduli of the D2-brane wrapped on these curves are the ones coming from the gauge field \textit{i.e.}, the moduli space of the D-brane wrapped on $(k+1)E$ is $T^{2}$,
\bea
N_{(k+1)E}^{j_{L},j_{R}}=\delta_{j_{R},0}\,\delta_{j_{L},\frac{1}{2}}\,.
\eea
Using these invariants in Eq. \ref{pf} we get $(a=1,2)$
\bea\nn
Z_{kE+C_{a}}&=&\prod_{i,j=1}^{\infty}(1-q_{1}^{i-\frac{1}{2}}\,q_{2}^{j-\frac{1}{2}}\,Q_{\bullet}^{k}Q_{a})\\\nn
Z_{(k+1)E}&=&\prod_{i,j=1}^{\infty}\Big[(1-q_{1}^{i}q_{2}^{j}Q_{\bullet}^{k+1})(1-q_{1}^{i-1}q_{2}^{j-1}Q_{\bullet}^{k+1})\Big]^{-1}\,.
\eea
The full partition function given by
\bea
Z=\prod_{k=0}^{\infty}\Big(Z_{kE+C_{1}}\,Z_{kE+C_{2}}\,Z_{(k+1)E}\Big)\,,
\eea
is exactly the same as Eq. \ref{cspf} for $(q_{1},q_{2})=(t^{-1},q^{-1})$.

If the preferred direction is chosen to be one of the internal line in the $X_{H}$ toric diagram \figref{Xh} then
the open string partition function has the form
\bea\label{ert}
Z_{\nu}=\sum_{\lambda}\,m(\lambda)\,f_{\nu}(\lambda)
\eea
Where $m(\lambda)$ is the probability measure on cylindric partitions with trivial profile and is given by
\bea
m(\lambda)=(Q_{1}Q_{2})^{|\lambda|}\prod_{s\in \lambda}\frac{(1-Q_{1}\,t^{a(s)+\frac{1}{2}}\,q^{\ell(s)+\frac{1}{2}})
(1-Q_{1}^{-1}\,q^{a(s)+\frac{1}{2}}\,t^{\ell(s)+\frac{1}{2}})}{(1-\,t^{a(s)+1}\,q^{\ell(s)})
(1-\,q^{\ell(s)+1}\,t^{a(s)})}
\eea
and
\bea f_{\nu}(\lambda)=s_{\nu^{t}}\Big(t^{\rho}\,q^{\lambda},Q_{1}\,t^{-\rho+\frac{1}{2}}\,q^{-\mu-\frac{1}{2}}\Big)
\eea
Thus we see that the open string partition function in this case is a correlation function in the Schur process. In terms of the gauge theory we know that the sum over $\lambda$ in Eq(\ref{ert}) is the sum over fixed points, under the $G=\mathbb{C}^{\times}\times \mathbb{C}^{\times}$ action, of the instanton moduli spaces which in this case are the Hilbert Scheme of $n$-points on $\mathbb{C}^2$. The effect of the insertion of the stack of branes is equivalent to insertion of an operator whose localization under $G$ gives $f_{\nu}(\lambda)$.  It will be interesting to identify these operators in the gauge theory although it is known in certain special cases that these are surface operators.

\subsection{Cylindric Partitions and Quantized K\"ahler Parameters}

In this section we will see that the open topological string partition function give by Eq(\ref{final}) acquires a combinatorial interpretation as the generating function of the cylindric partitions once we quantize the K\"ahler parameters $Q_{1}$ and $Q_{2}$. The quantization of the K\"ahler parameters appears naturally in the geometric transition which relates the Chern-Simons theory on the conifold with the topological string theory on the ${\cal O}(-1)\oplus {\cal O}(-1)\mapsto \mathbb{P}^{1}$ \cite{Gopakumar:1998ki}.

To see the effect of quantized K\"ahler parameters consider Eq(\ref{final}),
\bea\label{final2}
Z_{\nu}(Q_{1},Q_{2},t,q)&=&M(t,q)^{-1}\,q^{-\frac{||\nu||^2}{2}}\widetilde{Z}_{\nu}(t^{-1},q^{-1})\\\nn
&\times& \sum_{\lambda,\mu}\,Q_{\bullet}^{|\lambda|}\,
s_{\lambda/\mu}\Big(t^{\rho-\frac{1}{2}}\,q^{\nu+\frac{1}{2}},Q_{1}^{-1}t^{-\rho}\Big)\,
s_{\lambda/\mu}\Big(q^{\rho}\,t^{\nu^{t}},\sqrt{\frac{t}{q}}\,Q_{2}^{-1}q^{-\rho}\Big).
\eea
The arguments of the above skew-Schur functions are two infinite set of variables which reduce to finite sets when we specialize to certain values of $Q_{1}$ and $Q_{2}$ (Appendix B),
\bea\label{reduction}
t_{1}&=&\ell\,\epsilon_{2}+\Big(\frac{\epsilon_{1}+\epsilon_{2}}{2}\Big)\\\nn
t_{2}&=&-n\,\epsilon_{1}+\Big(\frac{\epsilon_{1}+\epsilon_{2}}{2}\Big)\\\nn
\{t^{\rho-\frac{1}{2}}\,q^{\nu+\frac{1}{2}},Q_{1}^{-1}t^{-\rho}\} &\xmapsto{Q_{1}=\sqrt{\frac{t}{q}}t^{\ell}}&\{t^{-i}\,q^{\nu_{i}+\frac{1}{2}}\,,\,i=1,\mathellipsis,\ell\}\\\nn
\{q^{\rho}\,t^{\nu^{t}},\sqrt{\frac{q}{t}}\,Q_{2}^{-1}q^{-\rho}\}&\xmapsto{Q_{2}=\sqrt{\frac{t}{q}}\,q^{n}}& \{q^{-j+\frac{1}{2}}\,t^{\nu^{t}_{j}}\,,\,j=1,\mathellipsis,n\},
\eea
with $\ell(\nu)\leq \ell\,,\,\ell(\nu^{t})\leq n$ and $Q_{\bullet}=Q_{1}Q_{2}=t^{\ell}\,q^{n}\left(\frac{t}{q}\right)$. Recall that $q=e^{\epsilon_{1}}\,,\,t=e^{-\epsilon_{2}}$ where $\epsilon_{1,2}$ are the Omega-background parameters \cite{nek}. After the above identification of the K\"{a}hler parameters Eq(\ref{final2}) takes the following form:
\bea
Z_{\nu}(Q_{1},Q_{2},t,q)&=&M(t,q)^{-1}\,q^{-\frac{\Arrowvert\nu\Arrowvert^2}{2}}\widetilde{Z}_{\nu}(t^{-1},q^{-1})\\\nn
&\times& \sum_{\lambda,\mu}\,Q_{\bullet}^{|\lambda|}\,
s_{\lambda/\mu}\Big(t^{-1}\,q^{\nu_{1}+\frac{1}{2}},\mathellipsis,t^{-\ell}\,q^{\nu_{\ell}+\frac{1}{2}}\Big)\,
s_{\lambda/\mu}\Big(q^{-\frac{1}{2}}\,t^{\nu^{t}_{1}},\mathellipsis,q^{-n+\frac{1}{2}}\,t^{\nu^{t}_{n}}\Big)\\\nn
&=&M(t,q)^{-1}\,q^{-\frac{\Arrowvert\nu\Arrowvert^2}{2}}\widetilde{Z}_{\nu}(t^{-1},q^{-1})\\\nn
&&\prod_{k=1}^{\infty}\Big((1-Q_{\bullet}^{k})^{-1}\prod_{i,j=1}^{\ell,n}(1-Q_{\bullet}^{k}\,t^{-i+\nu^{t}_{j}}\,q^{-j+\nu_{i}+1})^{-1}\Big),
\eea

According to the crystal picture of the topological vertex, the  open topological string partition function of $\mathbb{C}^3$ with three stack of branes on the three legs ($U(1)$ invariant locus) is the same as the generating function of the 3D partition (with boundaries specified by the stack of branes) after factoring out the MacMahon function \cite{ORV, Okuda:2004mb, sulkowski:2006jp,Iqbal:2008ra}. This factor is associated with the {\it vacuum} contribution \cite{ORV}. In other words, we need to multiply the partition function first with the MacMahon function before comparing it with the generating function of cylindric partitions. To this end, let us define
\bea
{\cal Z}^{\ell,n}_{\nu}(t,q)&=&M(t,q)\,Z_{\nu}(Q_{1},Q_{2},t,q)|_{Q_{1}=\sqrt{\frac{t}{q}}t^{\ell},Q_{2}=\sqrt{\frac{t}{q}}\,q^{n} }\\\nn
&=&q^{-\frac{\Arrowvert\nu\Arrowvert^2}{2}}\widetilde{Z}_{\nu}(t^{-1},q^{-1})\\\nn
&&\prod_{k=1}^{\infty}\Big((1-Q_{\bullet}^{k})^{-1}\prod_{i,j=1}^{\ell,n}(1-Q_{\bullet}^{k}\,t^{-i+\nu^{t}_{j}}\,q^{-j+\nu_{i}+1})^{-1}\Big).
\eea
In the previous section, we have computed the generating function for the cylindric partitions and a quick comparison shows that
\bea\nn\shadowbox{$
{\cal Z}^{\ell,n}_{\nu}(s^{-1},s^{-1})=\,s^{\frac{\Arrowvert\nu\Arrowvert^2}{2}}\,\mathbb{G}^{\ell,n}_{\nu}(s)$}
\eea

The more general partition function ${\cal Z}^{\ell,n}_{\nu}(t,q)$ can also be interpreted in terms of cylindric plane partitions. Given a cylindric plane partition $\pi$ let $\eta(a)$ be the 2D partitions obtained by diagonal slicing of $\pi$, $|\pi|=\sum_{a=1}^{N}|\eta(a)|$,
\bea
{\cal Z}^{\ell,n}_{\nu}(t,q)=q^{-\frac{\Arrowvert\nu\Arrowvert^2}{2}}\,\sum_{\substack{\mbox{\tiny cylindric partition $\pi$}\\ \mbox{\tiny of profile $\nu$} }}\,(t^{-1})^{\sum_{i,a_i=1}|\eta(i)|}\,(q^{-1})^{\sum_{j,b_j=1}|\eta(j)|}
\eea
Thus ${\cal Z}^{\ell,n}(t,q)$ is obtained if we count each slice of $\pi$ with parameter $t$ or $q$ depending on the shape of $\nu$.

\underline{\bf Level-Rank Duality}\,\,\,
The generating function $\mathbb{G}^{\ell,n}_{\nu}(s)$ has a symmetry:
\bea
\mathbb{G}^{\ell,n}_{\nu}(s)=\mathbb{G}^{n,\ell}_{\nu^t}(s).
\eea
This symmetry corresponds to the reflection of $n\times \ell$ rectangle across the diagonal. In \cite{tingley} it was shown that this symmetry is a manifestation of level-rank duality and the function $\mathbb{G}^{\ell,n}_{\nu}(s)$ is the character of a level $\ell$ irreducible representation of $\widehat{gl}_n$. In the topological string partition function this level-rank duality corresponds to the symmetry between the K\"ahler parameter $t_{1}$ and $t_{2}$, which in turn is due to the exchange symmetry between the two $\mathbb{P}^{1}$'s in the geometry.

\section{Conclusion}
In this paper we have shown that the random process, periodic Schur process, is closely related to the $N=2^{*}$ supersymemtric gauge theory as well as to the topological strings on a Calabi-Yau threefold. We showed that the partition function of this process is exactly the partition function of the $N=2^{*}$ supersymmetric gauge theory and moreover the partition function of the cyclindric partitions, on which the periodic Schur process is defined, with non-trivial profile is related to the open string partition function of the Calabi-Yau threefold which is used to engineer the $N=2^{*}$ $U(1)$ gauge theory. The profile of the cylindric partitions is given by the representation of the stack of branes placed on a toric 3-cycle in the Calabi-Yau threefold. These cylindric partitions appear to be labelling the fixed points of some quiver moduli space associated with this gauge theory under the lift of the $\mathbb{C}^{\times}\times \mathbb{C}^{\times}$ action on $\mathbb{C}^{2}$: $(z_{1},z_{2})\mapsto (q\,z_{1},t^{-1}\,z_{2})$. This is under investigation.

\section*{Appendix A: $X_{H}$ and Holomorphic Curves}

\begin{figure}[h]\begin{center}
\includegraphics[width=4in]{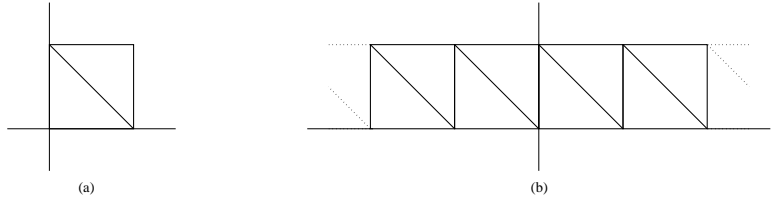}
\caption{\small The toric diagram of the resolved conifold and the infinite cover of the toric diagram of $X_{H}$.}  \label{fan}
\end{center}\end{figure}

In this section we will argue that the only holomorphic curves in the geometry $X_{H}$ are the ones given by Eq(\ref{holo}). The geometry ${\cal O}(-1)\oplus {\cal O}(-1)\mapsto \mathbb{P}^{1}$ is toric and has the Newton polygon shown in \figref{fan}(a). The geometry $X_{H}$ has a non-planar Newton polygon given by identifying the two vertical edges of the Newton polygon given in \figref{fan}(a). Thus for this geometry the Newton polygon lives on a cylinder. On the covering space of this cylinder we can represent the Newton polygon of $X_{H}$ by \figref{fan}(b).

\begin{figure}[h]\begin{center}
\includegraphics[width=4in]{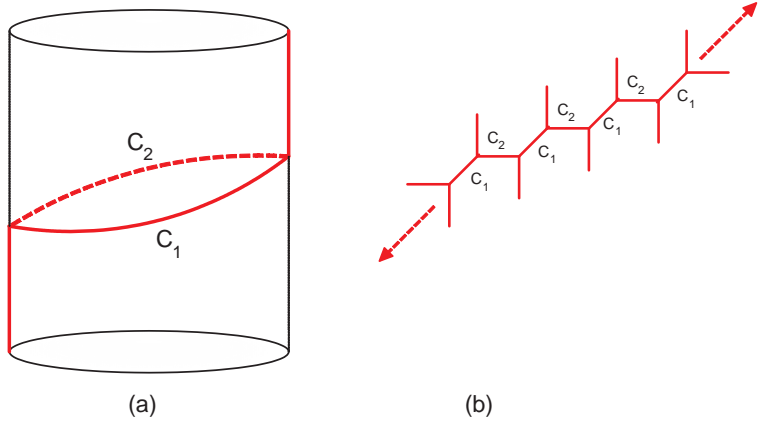}
\caption{\small The web diagram of $X_{H}$ (a) and its cover (b).}  \label{toric}
\end{center}\end{figure}

The web diagram corresponding to \figref{fan}(b) is shown in \figref{toric}(b).

The compact part of this geometry is an infinite chain of $\mathbb{P}^1$'s. Since this corresponds to the cover of $X_{H}$ there are only two distinct curves $C_{1}$ and $C_{2}$. The covering geometry is actually the maximal blowup of the resolution of $\mathbb{C}^{2}/\mathbb{Z}_{N}\times \mathbb{C}$ in the limit $N\mapsto \infty$. The holomorphic curves in this geometry satisfy $C\cdot C=2g-2$ for $g\geq 0$, where $g$ is the genus of the curve $C$. The only curves in this geometry are rational curves given by connected chains of $\mathbb{P}^1$'s. If $A$ and $B$ are two rational curve with $A\cdot B=0$ then $(A+B)\cdot (A+B)=-4$ hence $A+B$ is not holomorphic unless they form a connected chain. It is easy to see that connected chains belong to one of the following curve classes:
\bea
&&(k+1)(C_{1}+C_{2})\,,\,\,\,\mbox{If the first and last cycle in the chain are not the same,}\\\nn
&&k(C_{1}+C_{2})+C_{1}\,,\,\,\mbox{If the first and last cycle in the chain are both $C_{1}$,}\\\nn
&&k(C_{1}+C_{2})+C_{2}\,,\,\,\mbox{If the first and last cycle in the chain are both $C_{2}$}\,.
\eea

\section*{Appendix B: Refined Topological Vertex Computation}
\par{In this section, we want to show the details of the refined vertex computation and argue that the reduction of the arguments of the skew-Schur function happens in Eq. \ref{reduction}. According to the regular gluing rules of the refined topological vertex, the partition function takes the following form:}
\bea
Z_{\nu}(Q_{1},Q_{2},t,q)&=&\sum_{\lambda,\mu}(-Q_{1})^{|\lambda|}(-Q_{2})^{|\mu|}C_{\lambda\mu\nu}(t^{-1},q^{-1})
C_{\lambda^{t}\mu^{t}\emptyset}(q^{-1},t^{-1})\\
\nonumber
&=&q^{-\frac{||\nu||^2}{2}}\widetilde{Z}_{\nu}(t^{-1},q^{-1})\sum_{\lambda,\mu,\eta_{1},\eta_{2}}(-Q_{1})^{|\lambda|}(-Q_{2})^{|\mu|}
\left(\frac{t}{q}\right)^{\frac{|\eta_{1}|-|\eta_{2}|}{2}}
s_{\lambda^{t}/\eta_{1}}(t^{\rho}\,q^{\nu})s_{\mu/\eta_{1}}(t^{\nu^{t}}\,q^{\rho})\\\nn
&\times&s_{\lambda/\eta_{2}}(q^{\rho})s_{\mu^{t}/\eta_{2}}(t^{\rho})\\\nn
&=&q^{-\frac{||\nu||^2}{2}}\widetilde{Z}_{\nu}(t^{-1},q^{-1})\sum_{\lambda,\mu,\eta_{1},\eta_{2}}Q_{1}^{|\lambda|}Q_{2}^{|\mu|}
\left(\frac{t}{q}\right)^{\frac{|\eta_{1}|-|\eta_{2}|}{2}}
s_{\lambda^{t}/\eta_{1}}(t^{\rho}\,q^{\nu})s_{\mu/\eta_{1}}(t^{\nu^{t}}\,q^{\rho})\\\nn
&\times&s_{\lambda^{t}/\eta_{2}^{t}}(q^{-\rho})s_{\mu/\eta_{2}^{t}}(t^{-\rho}),
\eea
where we made use of $s_{\lambda/\mu}(q^{\rho})=(-1)^{|\lambda|-|\mu|}s_{\lambda^{t}/\mu^t}(q^{-\rho})$ in the second equality above. We first perform the sum over $\eta_{2}$:
\bea
\sum_{\eta_{2}}\Big(\frac{t}{q}\Big)^{-\frac{|\eta_{2}|}{2}}\,s_{\lambda^{t}/\eta_{2}}(q^{-\rho})s_{\mu/\eta_{2}}(t^{-\rho})&=&\Big(\frac{t}{q}\Big)^{-\frac{|\lambda|}{2}}\prod_{i,j=1}^{\infty}(1-t^{i}\,q^{j-1})\\\nn
&\times&\sum_{\tau}s_{\tau/\lambda^{t}}(t^{-\rho})s_{\tau/\mu}(q^{-\rho-1/2}t^{1/2}).
\eea
At this point we can sum over the partition $\mu$ to combine two skew-Schur functions into one:
\bea
\sum_{\mu}Q_{2}^{|\mu|}s_{\tau/\mu}(q^{-\rho-1/2}t^{1/2})s_{\mu/\eta_{1}}(t^{\nu^{t}}q^{\rho})=Q_{2}^{|\eta_{1}|}\,s_{\tau/\eta_{1}}(q^{-\rho-1/2}t^{1/2},Q_{2}\,t^{\nu^{t}}q^{\rho}).
\eea
A similar sum should be performed over $\lambda$ to get the other skew-Schur function in the final result Eq. \ref{final}. Let us now show how the infinite number of arguments of the Schur function reduce to a finite number of arguments when we specialize the K\"ahler parameters. First note that the Schur functions can be written in terms of power sums:
\bea
s_{\lambda}(x)=\sum_{\mu}m_{\lambda\mu}p_{\mu}(x),
\eea
where $p_{\mu}(x)=p_{\mu_{1}}(x)p_{\mu_{2}}(x)\mathellipsis$ with $p_{k}(x)$'s are the power sums,
\bea
p_{k}(x)=\sum_{i=1}^{\infty}x_{i}^{k}\,.
\eea
Hence, we can conclude that our assertion is correct for the Schur functions once we show it for the power sums. The generalization to skew Schur functions is straight forward. For the power sum symmetric functions
\bea
p_{k}\Big(t^{\rho-\frac{1}{2}}\,q^{\nu+\frac{1}{2}},Q_{1}^{-1}\,t^{-\rho}\Big)&=&\sum_{i=1}^{\infty}t^{-ki}\,q^{k(\nu_i+\frac{1}{2})}+
Q_{1}^{-k}\sum_{i=1}^{\infty}
t^{k(i-\frac{1}{2})}\,,\,\,\,\,\,Q_{1}=\sqrt{\frac{t}{q}}t^{\ell}\\\nn
&=&\sum_{i=1}^{\ell}\,t^{-ki}\,q^{k(\nu_i+\frac{1}{2})}=p_{k}(t^{-1}q^{\nu_{i}+\frac{1}{2}},t^{-2}q^{\nu_{2}+\frac{1}{2}},\cdots,t^{-\ell}q^{\nu_{\ell}+\frac{1}{2}})\\\nn
p_{k}(q^{\rho}\,t^{\nu^{t}},Q_{2}^{-1}\,q^{-\rho-\frac{1}{2}}t^{\frac{1}{2}})&=&\sum_{j=1}^{\infty}q^{k(-j+\frac{1}{2})}\,t^{k\nu_{j}^t}
+\sqrt{\frac{t^k}{q^k}}Q_{2}^{-k}\sum_{j=1}^{\infty}q^{k(i-\frac{1}{2})}\,,\,\,\,\,Q_{2}=\sqrt{\frac{t}{q}}q^{n}\\\nn
&=&\sum_{j=1}^{n}q^{k(-j+\frac{1}{2})}\,t^{k\nu_{j}^t}\,=p_{k}(q^{-\frac{1}{2}}t^{\nu_{1}^t},q^{-\frac{3}{2}}t^{\nu_{2}^t},\cdots,q^{-n+\frac{1}{2}}t^{\nu_{n}^t})\,,
\eea
where we have used analytic continuation to write the infinite sum over $t^{i}$ and $q^{j}$ in terms of $t^{-1}$ and $q^{-1}$.

\end{document}